\documentclass[prb,
 superscriptaddress, twocolumn,
aps]{revtex4-1} 

\usepackage{graphicx}
\usepackage{hyperref}
\usepackage{amsmath}
\usepackage{url}
\usepackage{color}
\usepackage{verbatim}
\usepackage{gensymb}
\usepackage{hyperref}
\usepackage{amssymb}
\usepackage{bm}

\def\beq {\begin{equation}}
\def\eeq {\end{equation}}
\def\w {\omega}
\def\bfq {\mathbf{q}}
\def\bfqr {\mathbf{q}_{r}}

\def\bfG {\mathbf{G}}
\def\bfk {\mathbf{k}}
\def\bfr {\mathbf{r}}

\newcommand{\soleil}{Synchrotron SOLEIL, L'Orme des Merisiers, Saint-Aubin, BP 48, F-91192 Gif-sur-Yvette, France}
\newcommand{\lsi}{LSI, CNRS, CEA/DRF/IRAMIS, \'Ecole Polytechnique, Institut Polytechnique de Paris, F-91120 Palaiseau, France}

\newcommand{\etsf}{European Theoretical Spectroscopy Facility (ETSF)}
\newcommand{\esrf}{European Synchrotron Radiation Facility, 71 Avenue des Martyrs, 38043 Grenoble, France}

\newcommand{\sorbonne}{Laboratoire de Chimie Physique – Mati\`ere et Rayonnement, Sorbonne Universit\'e, CNRS, F-75252 Paris, France}

\newcommand{\oxford}{Department of Physics, Oxford University, Clarendon Laboratory, Oxford, OX1 3PU, United Kingdom}

\begin{document}

\title{Dynamical screening in SrVO$_3$: Inelastic x-ray scattering experiments and ab initio calculations}

\author{Kari Ruotsalainen}
\email[]{Current affiliation Helmholtz-Zentrum Berlin, Germany, kari.ruotsalainen@helmholtz-berlin.de}

\affiliation{\soleil}
\affiliation{\lsi}

\author{Alessandro Nicolaou}
\affiliation{\soleil}

\author{Christoph J. Sahle}
\affiliation{\esrf}

\author{Anna Efimenko}
\affiliation{\esrf}

\author{James M. Ablett}
\affiliation{\soleil}

\author{Jean-Pascal Rueff}
\affiliation{\soleil}
\affiliation{\sorbonne}

\author{Dharmalingam Prabhakaran}
\affiliation{\oxford}

\author{Matteo Gatti}
\affiliation{\lsi}
\affiliation{\etsf}
\affiliation{\soleil}

\date{\today}

\begin{abstract}
We characterize experimentally and theoretically the high-energy dielectric screening properties of the prototypical correlated metal SrVO$_3$. The dynamical structure factor measured by inelastic x-ray scattering spectroscopy as a function of momentum transfer is in very good agreement with first-principles calculations  in the adiabatic local density approximation to time-dependent density-functional theory. Our results reveal the crucial importance of crystal local fields in the charge response function of correlated materials: They lead to depolarization effects for localised excitations and couple spectra from different Brillouin zones.
\end{abstract}

\maketitle

\section{Introduction}

SrVO$_3$ is a cubic perovskite with a rather simple electronic structure. The V ions are in a $d^1$ electronic configuration in an octahedral environment with three  $t_{2g}$ bands, well separated from other bands, crossing the Fermi level. It is hence 
easily amenable to a description based on the Hubbard model and  
has been long considered as a  prototype  strongly  correlated  metal \cite{Fujimori1992,Martin2016}. Indeed, despite experimental issues related to the surface sensitivity of early photoemission measurements\cite{Fujimori1992,Inoue1995,Sekiyama2004,Maiti2006,Eguchi2006,Laverock2013} and the formation of oxygen vacancies\cite{Backes2016}, in the last fifteen years SrVO$_3$ has served as a testbed  for the development of dynamical mean-field theory (DMFT) implementations for real materials \cite{Pavarini2004,Sekiyama2004,Nekrasov2005,Lechermann2006,Amadon2008,Huang2012,Tomczak2012,Lee2012,Sakuma2013,Taranto2013,Nilsson2017}. 

The dynamical screening of the Coulomb interaction, evaluated in the random-phase approximation (RPA), is the key physical ingredient that within Hedin's GW approximation \cite{Hedin1965} describes electronic correlation  beyond the Hartree-Fock approximation. 
The importance of dielectric screening in order to renormalize the electron-electron interaction has also been recognised 
for using the Hubbard model in solids\cite{Anisimov1997,Aryasetiawan2004}. 
In the recent past, the frequency dependent RPA screening has been often employed within the DMFT framework
to evaluate the on-site Hubbard interaction U in SrVO$_3$ \cite{Aryasetiawan2004,Aryasetiawan2006,Miyake2008,Vaugier2012,Casula2012,Nomura2012}. 
On the other hand, on the basis of {\it ab initio} GW calculations\cite{Gatti2013}, it was recently demonstrated that the coupling with low-energy plasmon excitations 
is responsible for the photoemission satellites  of SrVO$_3$. These results (see also Ref. \cite{Nakamura2016}) called for a reconsideration of correlation effects in SrVO$_3$, which was later confirmed by extended DMFT calculations as well \cite{Boehnke2016,Nilsson2017,Petocchi2020}. 

Both recent DMFT and GW calculations therefore indicate the crucial importance of determining accurately the dielectric screening properties of SrVO$_3$. Nevertheless, to the best of our knowledge, a detailed experimental investigation is still lacking \footnote{Resonant inelastic X-ray scattering (RIXS) has recently been used to study SrVO$_3$ (see Ref. \cite{McNally2019}), but these experiments do not probe the dielectric function}.
Nonresonant inelastic X-ray scattering  (IXS) experiments measure the dynamic structure factor $S(\bfq,\w)$ and can therefore provide such information \cite{Schulke2007}. Thanks to IXS it is possible to have access to both plasmons (i.e., collective charge excitations) and various kinds of interband transitions, together with their dispersion as a function of the wavevector $\bfq$. Furthermore, IXS experiments use hard X-rays resulting in true bulk sensitivity and reduced sample damage per unit volume.
On the theoretical side, the dynamic structure factor 
can be calculated within {\it ab initio} linear-response time-dependent density-functional theory \cite{Runge1984,Onida2002,Ullrich2012} (TDDFT)  and directly compared with experiment.

In the present article we therefore tackle this question.  
By combining IXS experiments and first-principles TDDFT calculations, we present  a detailed investigation  of the high-energy electronic excitations 
of SrVO$_3$ as a function of the momentum transfer $\bfq$.
An accurate knowledge of those excitations also informs the construction of low-energy models within DMFT\cite{Aryasetiawan2004}: they are the charge fluctuations that are used to calculate the Hubbard $U$\footnote{We note that, contrary to the dynamic structure factor $S(\bfq,\w)$, the onsite interaction $U$  is an auxiliary effective parameter and not a directly measurable quantity. Indeed, its value depends on the choice of the one-particle orbitals in the Hubbard model hamiltonian. Moreover, different definitions and recipes have been proposed in literature for its calculation, see e.g. Refs. \cite{Anisimov1991,Aryasetiawan2004, Cococcioni2005}.}.

In principle, TDDFT is an exact method to describe neutral electronic excitations in materials. However, in practice 
{\it ab initio} approaches have to adopt approximations such as the adiabatic local density approximation\cite{Zangwill1980} (ALDA). 
They have been shown to provide reliable loss spectra for $sp$ semiconductors and metals\cite{Onida2002,Botti2007}, but have rarely been used for strongly correlated $df$ compounds (see e.g. Refs. \cite{Larson2007,Lee2010,Roedl2012,Huotari2010,Iori2012,Roedl2017}).
Our detailed comparison of TDDFT simulations and experimental results therefore provides an important benchmark to establish the  extent to which most widely used TDDFT approximations can be useful for describing and understanding the physics of electronic screening also in correlated metals\footnote{For  TDDFT applications in Hubbard models see e.g. Refs. \onlinecite{Verdozzi2008,Karlsson2011}.}.

The article is organised as follows.
In Sec. \ref{methods} we outline the experimental and computational details. 
In Sec. \ref{results}, on the basis of the very good agreement between measured and calculated spectra, we analyse the origin of main spectral features and we unravel the strong impact of the induced crystal local fields, which turn out to be an essential aspect of the dielectric response of SrVO$_3$.
Finally, Sec. \ref{conclusions} summarizes our work.

\section{Methods}
\label{methods}

\subsection{Experiment}

The experiments were performed at beamline ID20 of the European Synchrotron Radiation Facility. The data were acquired using the RIXS\cite{MorettiSala2018} and Raman\cite{Huotari2017} spectrometers. We used an incident photon energy of 6455.93 eV. The incident flux was monochromatized employing a Si(111) monochromator and Si(440) analyzer crystals with a 2 m bending radius were used. The total experimental energy resolution was approximatively 1 eV (FWHM) as read off the elastic line. 

The sample was a SrVO$_3$ single crystal grown by the floating zone technique. The orientation was studied with the Laue method and the crystal surface was prepared by polishing followed by removal of the damaged layers with nitric acid. With the RIXS spectrometer we used a $\theta$-$2\theta$ scattering geometry with $\textbf{q}$ oriented along the $(100)$ direction. With the Raman spectrometer we used a multi-analyzer setup and the momentum transfers have components along a perpendicular reciprocal lattice vector as well. The momentum resolution was of the order of 10\% of $\bfG=(100)$ at the low scattering angles and less than 1\% at the highest.  The experiments were performed at room temperature.  
The tail of the elastic peak gives uncertainty to the experimental results for $\w \lesssim 5$ eV and, in the following, the spectra are reported and discussed in the energy range of 5-50 eV.
Therefore, the investigation of $dd$ excitations at energies below 5 eV is left for a future experimental work.

\subsection{Theory}

The dynamic structure factor $S(\bfq,\w)$ for a transferred momentum\footnote{If not otherwise specified, atomic units are used throughout the paper.}  $\bfq$ describes the longitudinal response of an electronic system to an external scalar potential.
 Thanks to the fluctuation-dissipation theorem,  it can be related to the loss function $-\text{Im} \epsilon_{M}^{-1}$ as
 ($n$ is the mean electron density)\cite{Schulke2007}:
\begin{align}
S(\bfq,\w) =  &  -\frac{q^2}{4\pi^2n} \text{Im} \epsilon_{M}^{-1}(\bfq,\w) \nonumber \\
=  &  \frac{q^2}{4\pi^2n} \frac{\text{Im} \epsilon_{M}(\bfq,\w)}{\left[\text{Re} \epsilon_{M}(\bfq,\w)\right]^2+\left[\text{Im} \epsilon_{M}(\bfq,\w)\right]^2}.
\label{eq:sqw}
\end{align}
Plasmon resonances  correspond to the zeros of $\text{Re} \epsilon_{M}$ for $\text{Im} \epsilon_{M}$ not too large, while spectral features associated to (screened) interband transitions can be detected as peaks of $\text{Im} \epsilon_{M}$ as well. 
$\epsilon_{M}$ is inversely proportional to a diagonal element of the inverse dielectric matrix\cite{Adler1962,Wiser1963}:
\beq
\epsilon_{M}(\bfq=\bfqr + \bfG,\w) = \frac{1}{\epsilon^{-1}_{\bfG,\bfG}(\bfqr,\w)},
\label{micro:macro}
\eeq
where $\bfqr$ belongs to the first Brillouin zone and $\bfG$ is a reciprocal-lattice vector. For  $\bfq\rightarrow0$, the expression \eqref{micro:macro} yields the macroscopic dielectric function  $\epsilon_{M}(\bfq\rightarrow0,\w)$ that describes the optical spectra. In any inhomogeneous material, the off-diagonal elements $\bfG\neq\bfG'$ of the dielectric matrix $\epsilon_{\bfG,\bfG'}$ are not zero and therefore contribute to the diagonal element of its inverse $\epsilon^{-1}_{\bfG,\bfG}$. 
By defining $M$ as the submatrix of $\epsilon$ that does not have the row and column $\bfG$, Eq. \eqref{micro:macro} can be rewritten as:
\begin{multline}
\epsilon_{M}(\bfq=\bfqr + \bfG,\w) = \epsilon_{\bfG,\bfG}(\bfqr,\w)  \\ -  \sum_{\bfG_1,\bfG_2\neq\bfG} \epsilon_{\bfG,\bfG_1}(\bfqr,\w) M^{-1}_{\bfG_1,\bfG_2}(\bfqr,\w)\epsilon_{\bfG_2,\bfG}(\bfqr,\w).
\label{lfe}
\end{multline}
The off-diagonal contributions due to inhomogeneities in the response functions explicitly appear in the second row of Eq. \eqref{lfe}: they are called crystal local-field effects (LFE).

The inverse dielectric function is related to the density-density response function $\chi$ via the Coulomb interaction  $v_c$:
\beq
\epsilon^{-1}_{\bfG,\bfG'}(\bfqr,\w) = \delta_{\bfG,\bfG'} +  v_c(\bfqr+\bfG)\chi_{\bfG,\bfG'}(\bfqr,\w).
\eeq
Within TDDFT  $\chi$ can be obtained as a solution of the Dyson-like equation \cite{Petersilka1996}:
\begin{multline}
\chi_{\bfG,\bfG'}(\bfqr,\w) = \chi^0_{\bfG,\bfG'}(\bfqr,\w)  +  \sum_{\bfG_1,\bfG_2}\chi^0_{\bfG,\bfG_1}(\bfqr,\w) \times \\
\left[v_{c}(\bfqr+\bfG_1)\delta_{\bfG_1,\bfG_2} + f^{xc}_{\bfG_1,\bfG_2}(\bfqr,\w)\right] \chi_{\bfG_2,\bfG'}(\bfqr,\w),
\label{dys-tddft}
\end{multline}
where $\chi^0$ is the Kohn-Sham (KS) independent-particle response function  and $f_{xc}$ is the exchange-correlation kernel.
The RPA corresponds to setting $f_{xc}=0$ in Eq. \eqref{dys-tddft}.
The simplest approximation beyond the RPA is the 
ALDA, where a local and static approximation to $f_{xc}$ is adopted.
$\chi^0$ is built from the KS\cite{Kohn1965} orbitals $\varphi_{n\bfk}$ and eigenvalues $\varepsilon_{n\bfk}$:
\begin{multline} 
\chi^0_{\bfG,\bfG'}(\bfqr,\w) = \frac{1}{\Omega}\sum_{n,n',\bfk} (f_{n\bfk}-f_{n'\bfk+\bfqr}) \\ 
\times \frac{ \tilde{\rho}_{nn'\bfk}(\bfqr,\bfG)\tilde{\rho}^*_{nn'\bfk}(\bfqr,\bfG')} {\w+\varepsilon_{n\bfk}-\varepsilon_{n'\bfk+\bfqr} +i\eta } 
\label{eq:chi0}
\end{multline}
where $f_{n\bfk}$ are the Fermi occupation numbers, $\Omega$ the crystal volume, and $\eta \rightarrow 0^+$ gives rise to a Lorentzian broadening.
 The matrix elements $\tilde{\rho}_{nn'\bfk}(\bfqr,\bfG)= \int d\bfr \varphi^*_{n\bfk}(\bfr) e^{-i(\bfqr+\bfG)\cdot\bfr}\varphi_{n'\bfk+\bfqr}(\bfr) $
give the oscillator strengths.

\begin{figure*}[!ht]
\includegraphics{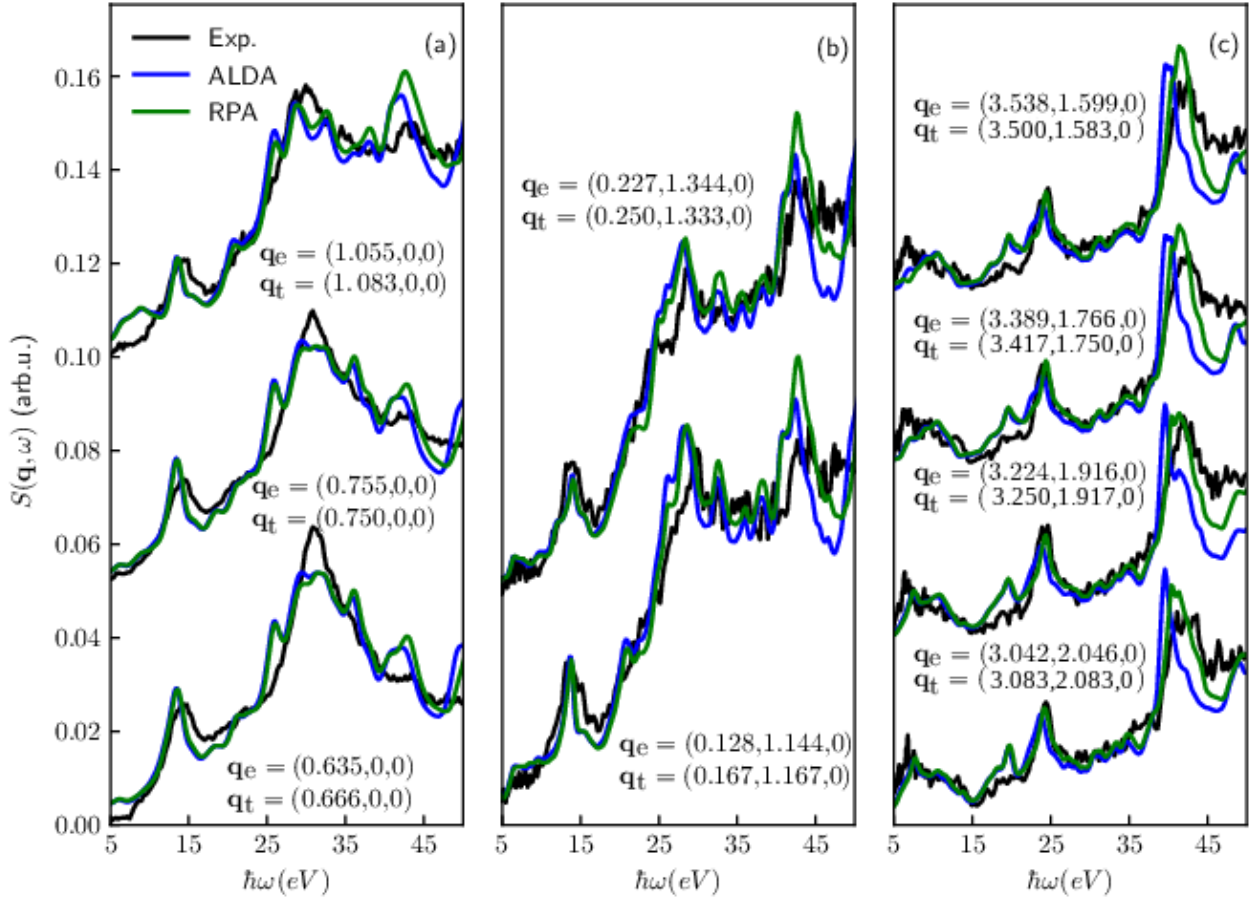}
\caption{ Comparison between experimental IXS spectra (black) and theoretical dynamic structure factors calculated within the ALDA (blue) and the RPA (green). $\bf{q}_\textrm{e(t)}$ denotes the experimental (theoretical) momentum transfer value in reciprocal lattice units (r.l.u.) and $\hbar\omega$ is the energy transfer ($\bf{q}_\textrm{t}$  is the wavevector in the 12x12x12 grid that is the closest to $\bf{q}_\textrm{e}$).
The experimental and theoretical spectra have been normalized to the area under the curve between 5 eV and 40 eV. A vertical offset has been added to the spectra for improved clarity.
(a) Low $q$. (b) Intermediate $q$. (c) High $q$.  
The theoretical spectra have been broadened by 0.17 (0.4) eV for panel (a) [(b),(c)]) to match the experimental resolution.  
 } 
\label{fig1}
\end{figure*}

In the RPA and when LFE are neglected,  $\textrm{Im} \epsilon_M$ reduces to the independent-particle picture of a sum over transitions $(n\bfk) \rightarrow (n'\bfk+\bfqr)$:
\begin{multline} 
\textrm{Im} \epsilon_M(\bfq,\w) =   \frac{4\pi^2}{\Omega q^2} \sum_{nn'\bfk} (f_{n\bfk}-f_{n'\bfk+\bfqr})\\ \times  \left| \tilde{\rho}_{nn'\bfk}(\bfqr,\bfG) \right|^2 \delta(\w -(\varepsilon_{n'\bfk+\bfqr} - \varepsilon_{n\bfk})),
\label{epsnlf}
\end{multline} 
whereas LFE and exchange-correlation effects beyond the RPA effectively mix all those electron-hole transitions.

In the present work, ground-state KS orbitals and energies are obtained in the local-density approximation (LDA) within a plane-wave basis approach on a 12$\times$12$\times$12 $\Gamma$-centered $\bfk$-point grid with norm-conserving pseudopotentials (including V $3s$ and $3p$ and Sr $4s$ and $4p$ semicore states) and an energy cutoff of 100 Hartree. To calculate $\chi^0$ from Eq. \eqref{eq:chi0}, 200 bands have been included in the sum and the matrix size has a 245 eV cutoff. We have adopted the experimental lattice parameter of the  cubic perovskite crystal structure \cite{Lan2003} $a=3.8425$ \AA. Spin-orbit coupling has a minor effect on the electronic structure of SrVO$_3$ and has been neglected.
We have used \texttt{Abinit} \cite{abinit} for the KS ground-state calculations and  \texttt{Dp} \cite{dp} for the TDDFT simulations of IXS spectra.

\section{Results and  analysis}
\label{results}

\subsection{Comparison between experiments and calculations}

The low $q$  experimental IXS spectra are presented in Fig. \ref{fig1}(a) alongside with comparisons to ALDA and RPA calculations. Their main characteristics are a peak at 14.2 eV, a shoulder around 24 eV, a strong peak at 31 eV and a further peak at 41 eV. The 14 eV peak loses relative intensity upon increasing the momentum transfer while the high energy edge of the peak increases in intensity with respect to the peak maximum. Intensity is suppressed in the valley between the 14.2 eV and 24 eV peaks, and the latter increases in intensity. The main peak at 31 eV shows broadening towards the low energy side with increasing $q$, suggesting contributions from dispersive excitations, or non-dispersive ones with a q dependent intensity. Finally, a non-dispersive peak emerges at 42.6 eV.    

In Fig. \ref{fig1}(b) we present intermediate $q$ spectra. The low energy edge of the 14.2 eV peak shows a slight suppression of intensity with increasing $q$. The 24 eV shoulder continues to develop into a clear peak. Near 31 eV, the intensity is suppressed and the maximum is found at lower energies. The 42.6 eV peak increases in intensity as does the continuum past the peak maximum. 

Finally in Fig. \ref{fig1}(c) we present the high $q$ spectra. One immediately observes that the peak at 31 eV has completely merged with the continuum, and that the 24 eV peak remains. Furthermore the 41 eV peak is now the most prominent feature of the spectrum. We also note that the intensity region from 5-15 eV has also developed a double peak structure that was not clearly visible at low or intermediate $q$. There is a sharp peak at 6.6 eV followed by a broad feature extending up to 14 eV. Here the lineshapes remain roughly constant upon increasing $q$ and the spectra overlay very well across the whole energy range.   

Overall the calculated spectra present a slight underestimation of some peak positions and are more structured than the experimental spectra.  The inclusion of lifetime effects would damp many features, especially at high energies, and could further improve the agreement with experiment \cite{Weissker2006,Weissker2010,Cazzaniga2011,Huotari2011,Seidu2018}.
 Anyway, considering the fact that in the comparison  between calculations and measurements no adjustable parameters have been used, the agreement between ALDA and experiment is remarkable even for this supposedly strongly correlated transition-metal oxide.
 Furthermore, also RPA and ALDA results are qualitatively similar,  supporting the use of the RPA for the calculation\cite{Aryasetiawan2004} of the Hubbard $U$ in SrVO$_3$ \footnote{For an assessment of the constrained RPA downfolding procedure in model Hamiltonians we refer e.g. to Refs. \cite{Shinaoka2015,Honerkamp2018}.}.

\subsection{Dispersion}

\begin{figure}[t]
\includegraphics[angle=270,width=0.9\columnwidth,clip]{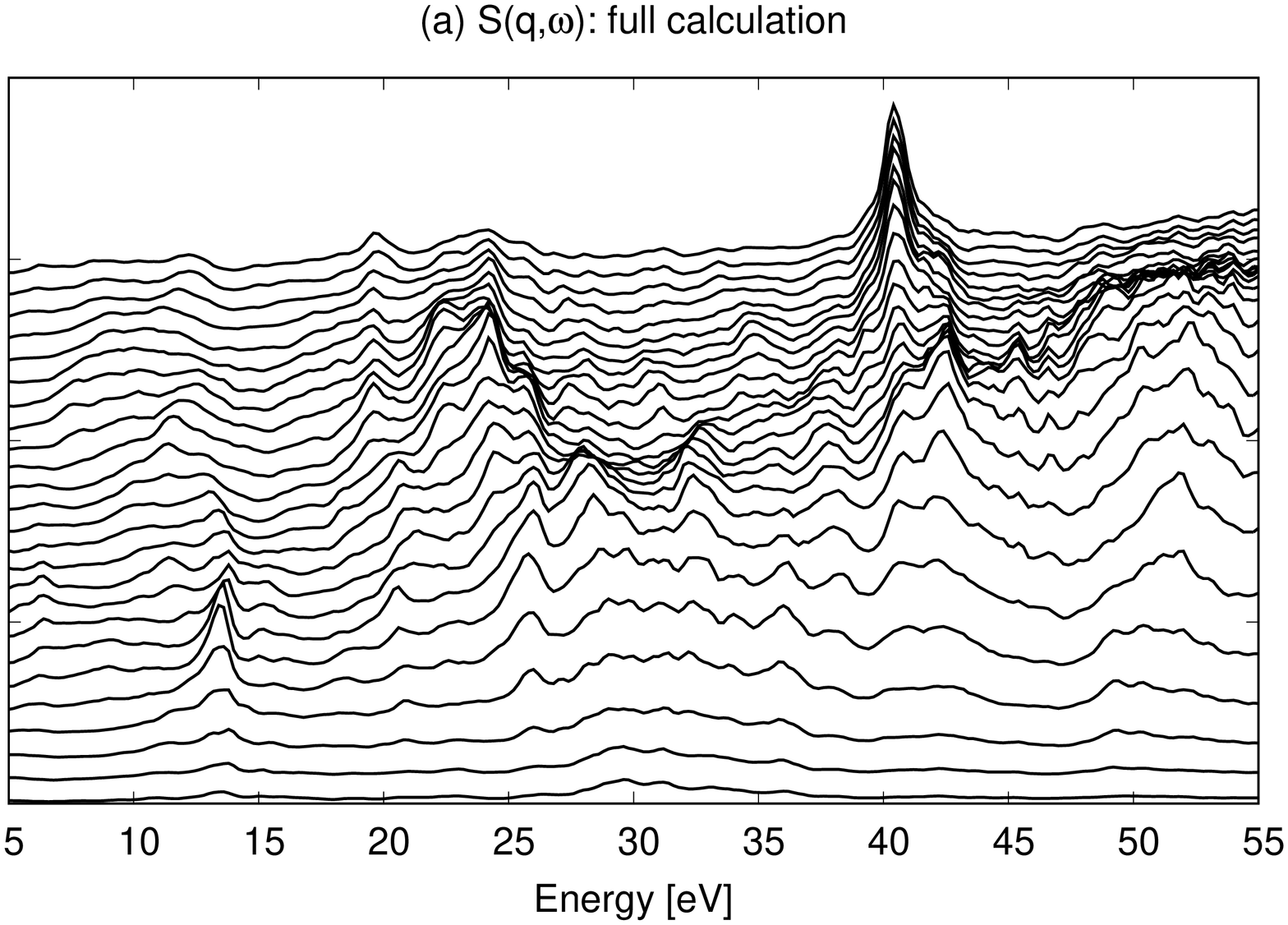}
\includegraphics[angle=270,width=0.9\columnwidth,clip]{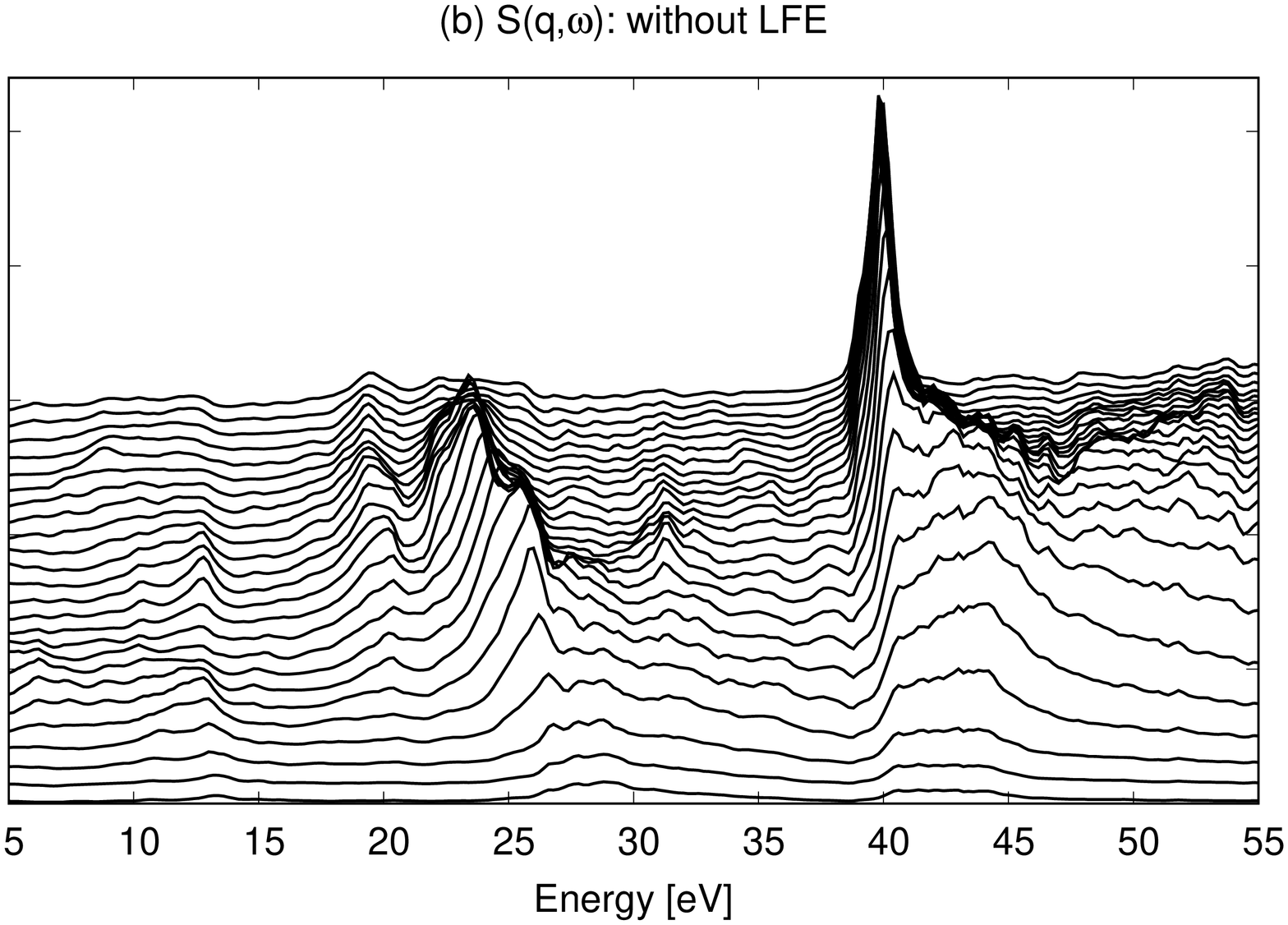}
\caption{Dispersion of the calculated dynamic structure factor $S(\bfq,\w)$ for $\bfq$ along the (100) direction. The size of $\bfq$ ranges from q=0.083 (bottom curve) to q=4.083  
(top curve) by steps equal to 0.166 
reciprocal lattice units. (a) Full calculations within the ALDA. (b) Calculations within the RPA neglecting local field effects (LFE).} 
\label{fig2}
\end{figure}

Having  verified  that  the  calculated screening captures the most important electronic excitations occurring in the measured loss function  with very good accuracy, we can safely use the simulated spectra to analyse more in detail the dispersion of the measured electronic excitations as a function of $\bfq$.

Fig. \ref{fig2}(a) shows the ALDA dynamic structure factor $S(\bfq,\w)$ for $\bfq$ along the $(100)$ direction and size $q$ 
spanning from $(1/12)G$ to $(49/12)G$, i.e. from 0.136 to 6.677 \AA$^{-1}$, with $(1/6)G$ steps.
Each curve, corresponding to increasing $q$ from the bottom to the top of the figure, has a vertical offset proportional to $q$.  In other nonequivalent directions the results are qualitatively similar and therefore not reported here for brevity.

The spectra are characterised by a rich set of features that do not display  large dispersions with $q$. Being nondispersive in reciprocal space means that they correspond to electronic excitations that are quite localised in real space. 
The shape of the spectra, instead, changes very much. Besides the gradual development of a broad continuum at large energies and  momentum transfers that is typical of Compton lineshapes\cite{Huotari2011,Ruotsalainen2015}, the relative spectral feature intensities are strongly varying  with $q$. Some peaks that are weak at small $q$ increase to become the most prominent features at large $q$.
The emergence of these peaks is even more noticeable in the spectra of Fig. \ref{fig2}(b) that are obtained neglecting entirely the LFE, i.e., from the first row of Eq. \eqref{lfe} calculated within the RPA.
Without LFE the spectra look very different, even qualitatively. Indeed, LFE turn out to be the crucial ingredient to reach a good agreement with experiment. 
A similar strong impact of crystal local fields was found in other materials where the electrons are analogously localized (see e.g. \cite{Gurtubay2004,Gurtubay2005,Iori2012,Gatti2015,Zhu2018}).
By comparing Fig. \ref{fig2}(b) and Fig. \ref{fig2}(a) we see that LFE mainly damp the most intense peaks (located  at $\sim$ 23 and 40 eV for large $q$) and redistribute their spectral weight to higher energies.

\begin{figure*}[t]
\includegraphics[angle=270,width=0.32\textwidth,clip]{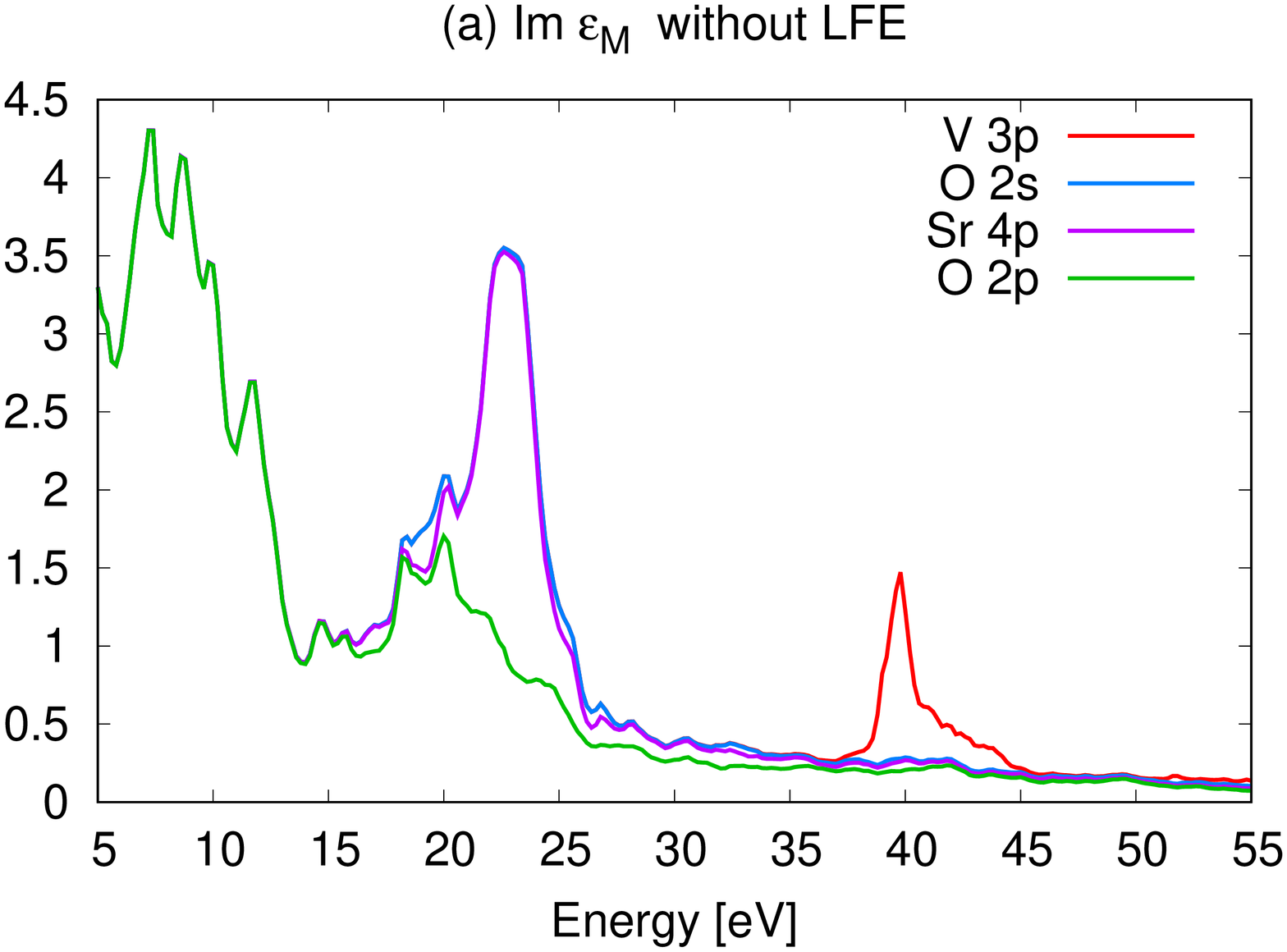}
\includegraphics[angle=270,width=0.32\textwidth,clip]{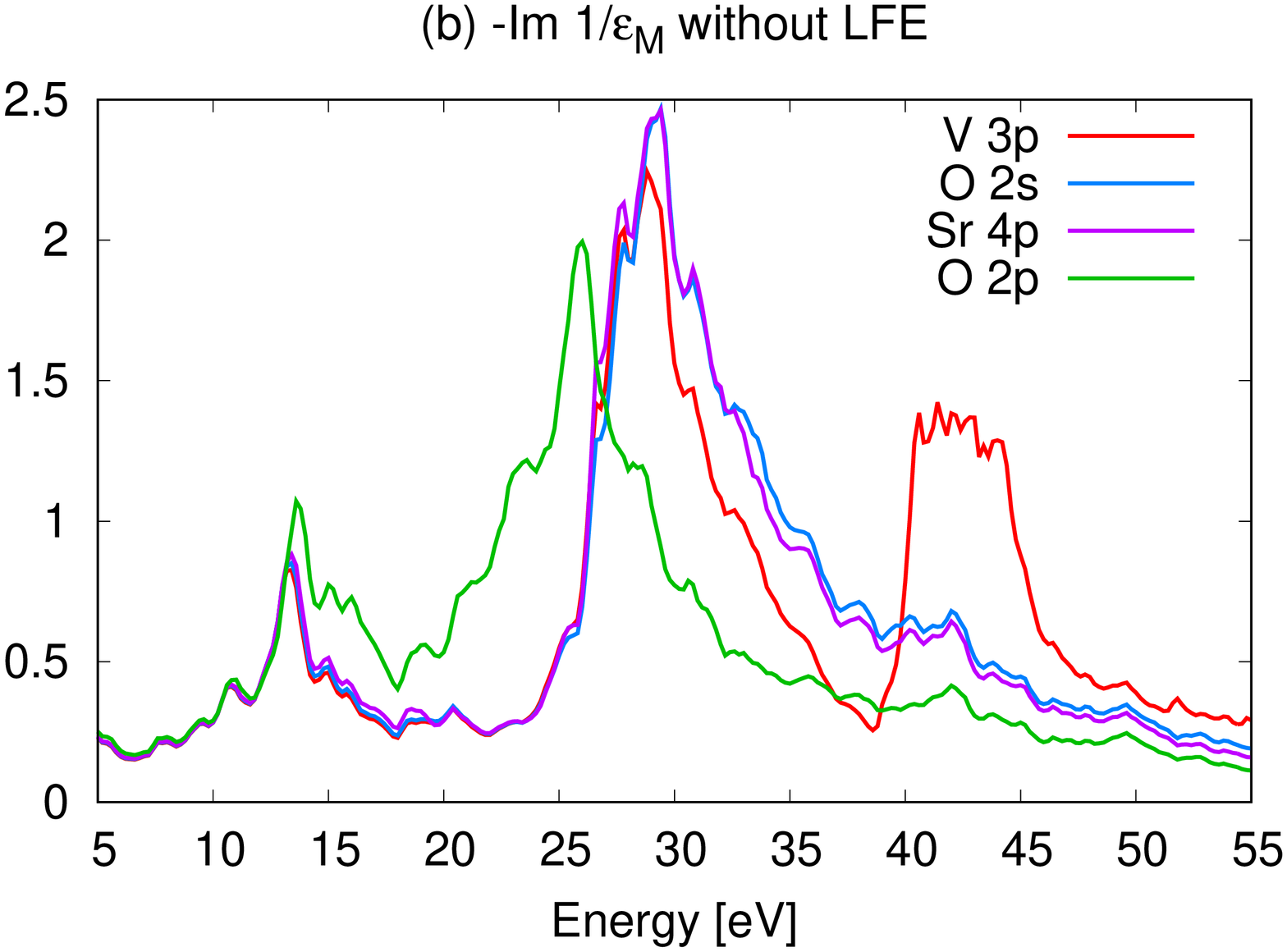}
\includegraphics[angle=270,width=0.32\textwidth,clip]{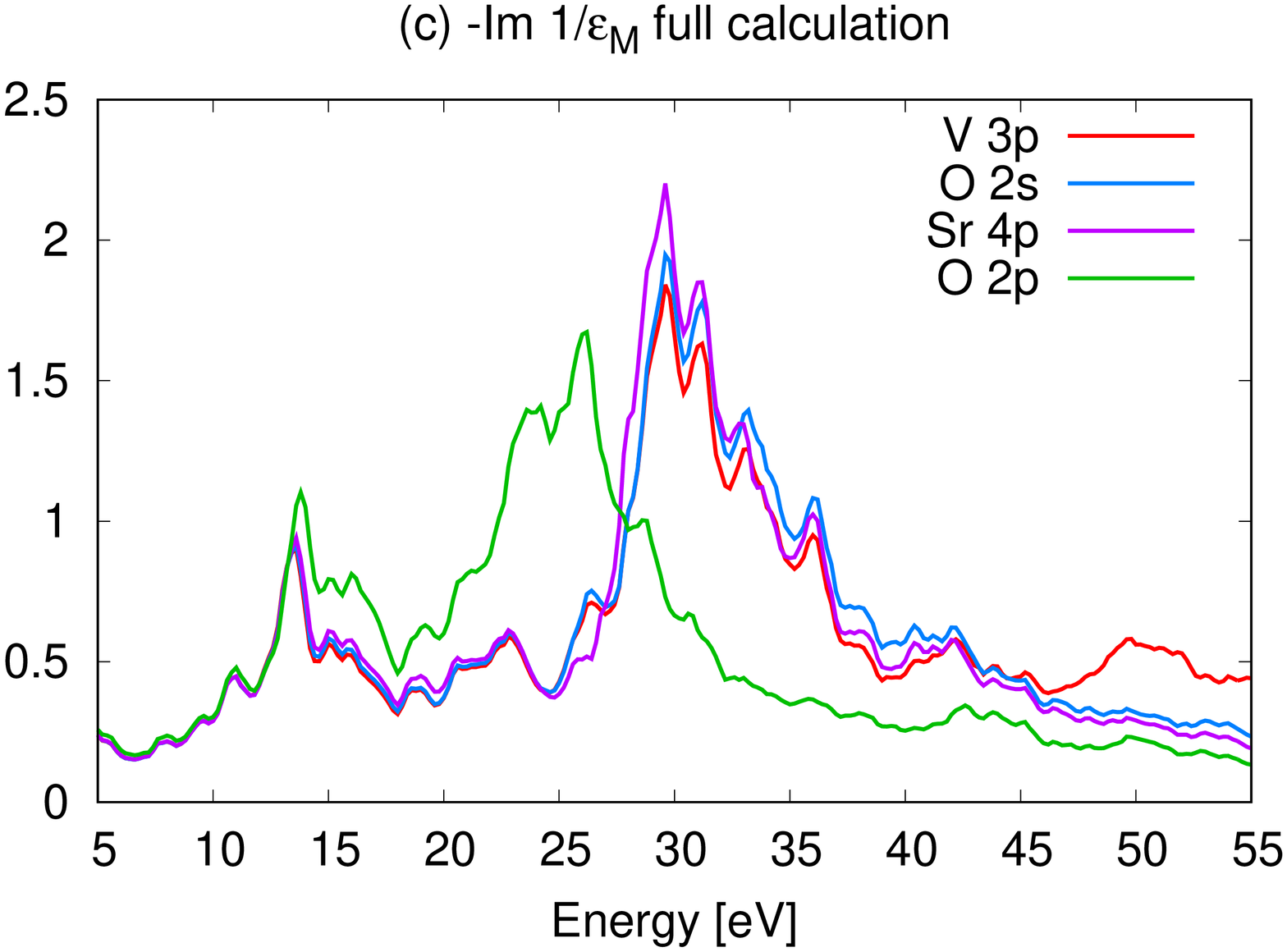} \\
\includegraphics[angle=270,width=0.32\textwidth,clip]{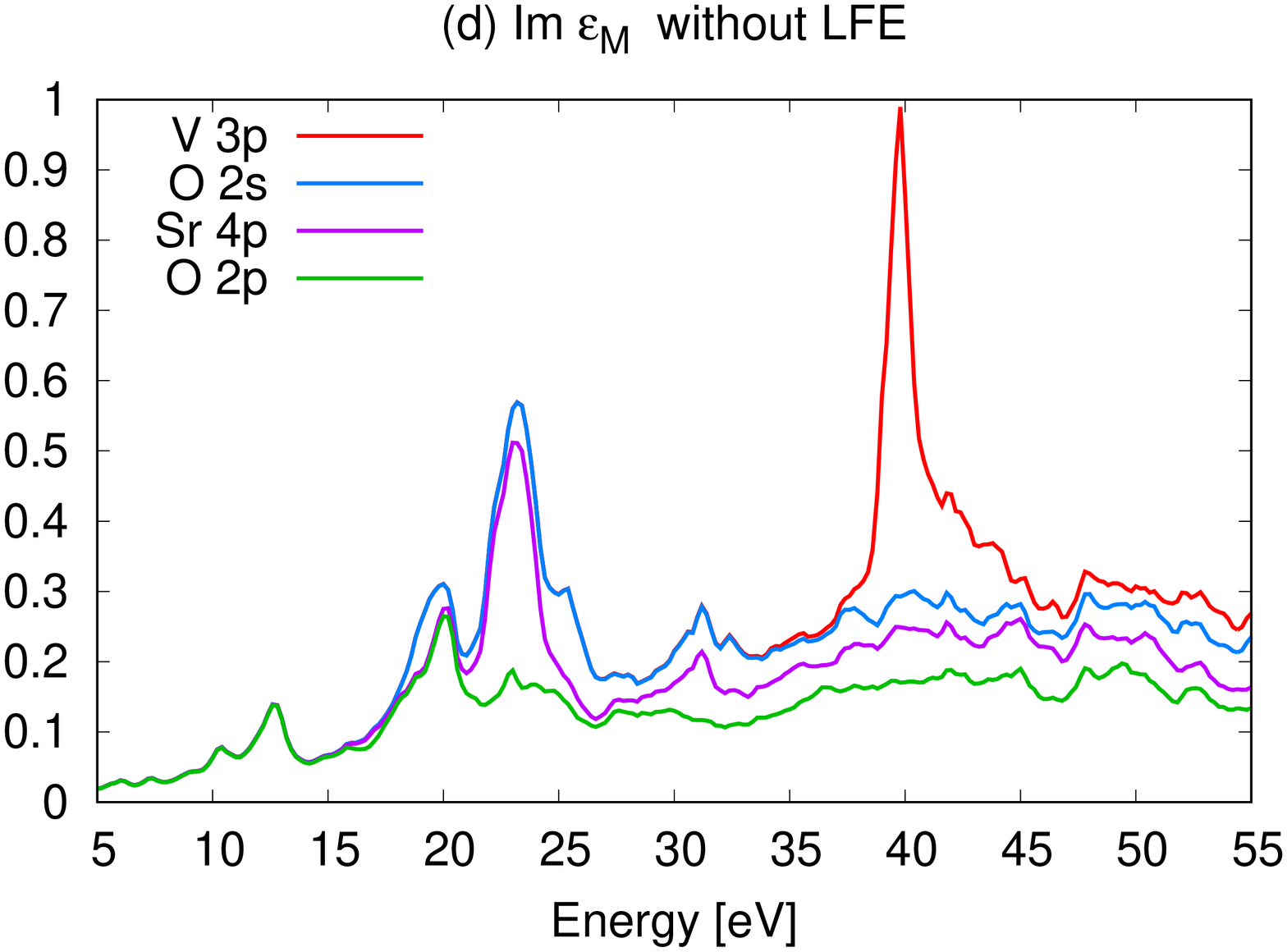}
\includegraphics[angle=270,width=0.32\textwidth,clip]{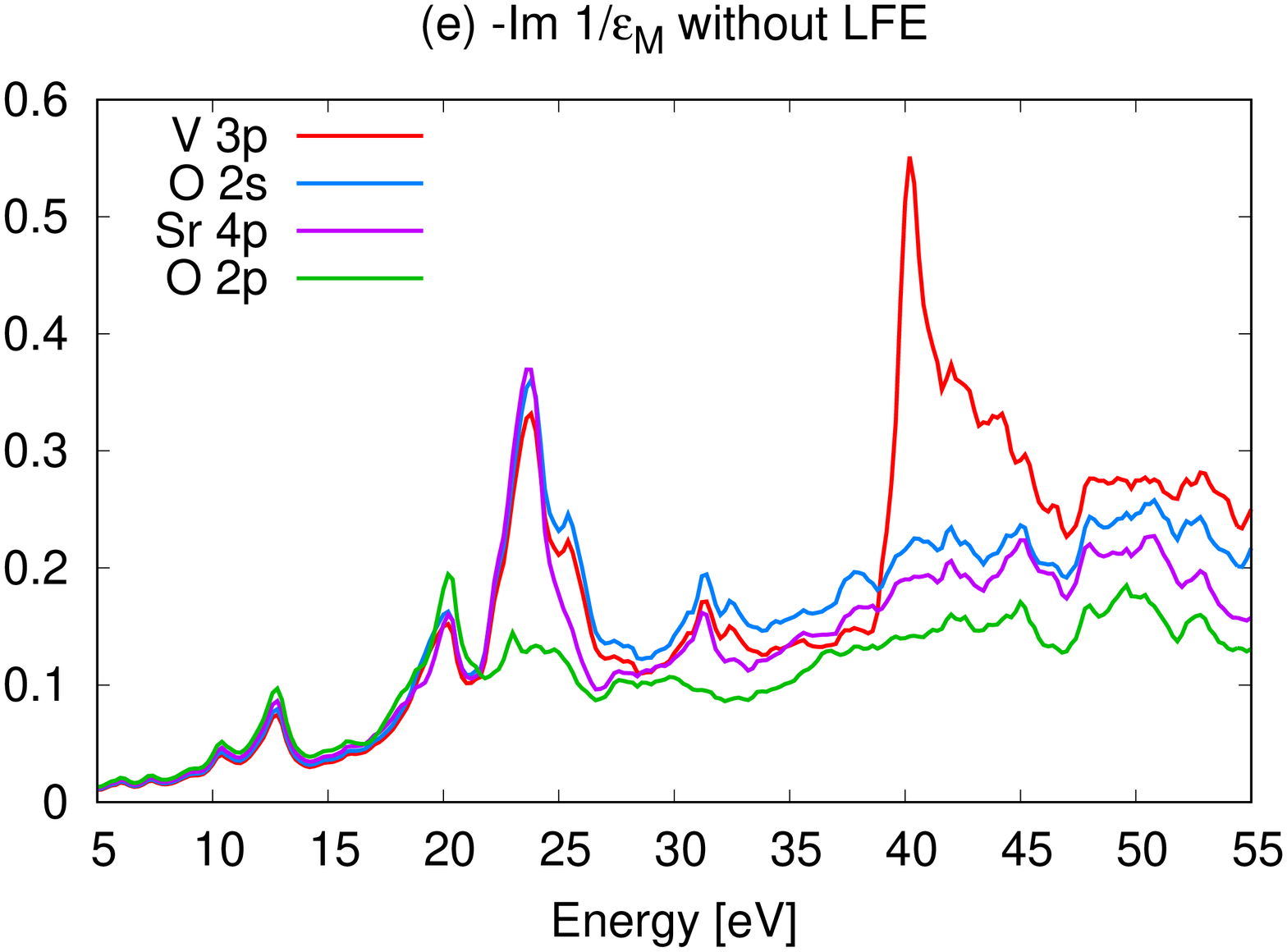}
\includegraphics[angle=270,width=0.32\textwidth,clip]{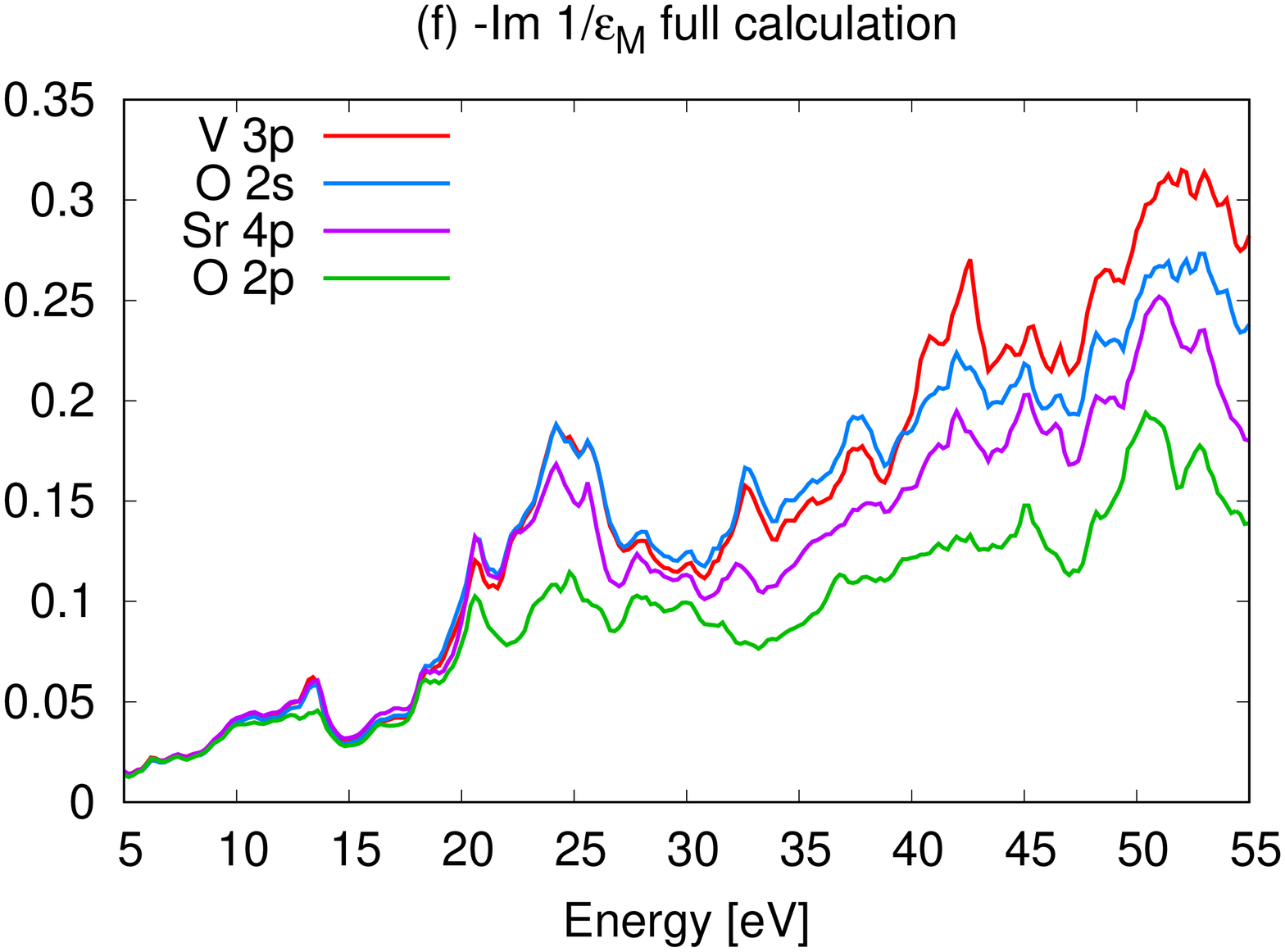}\\
\includegraphics[angle=270,width=0.32\textwidth,clip]{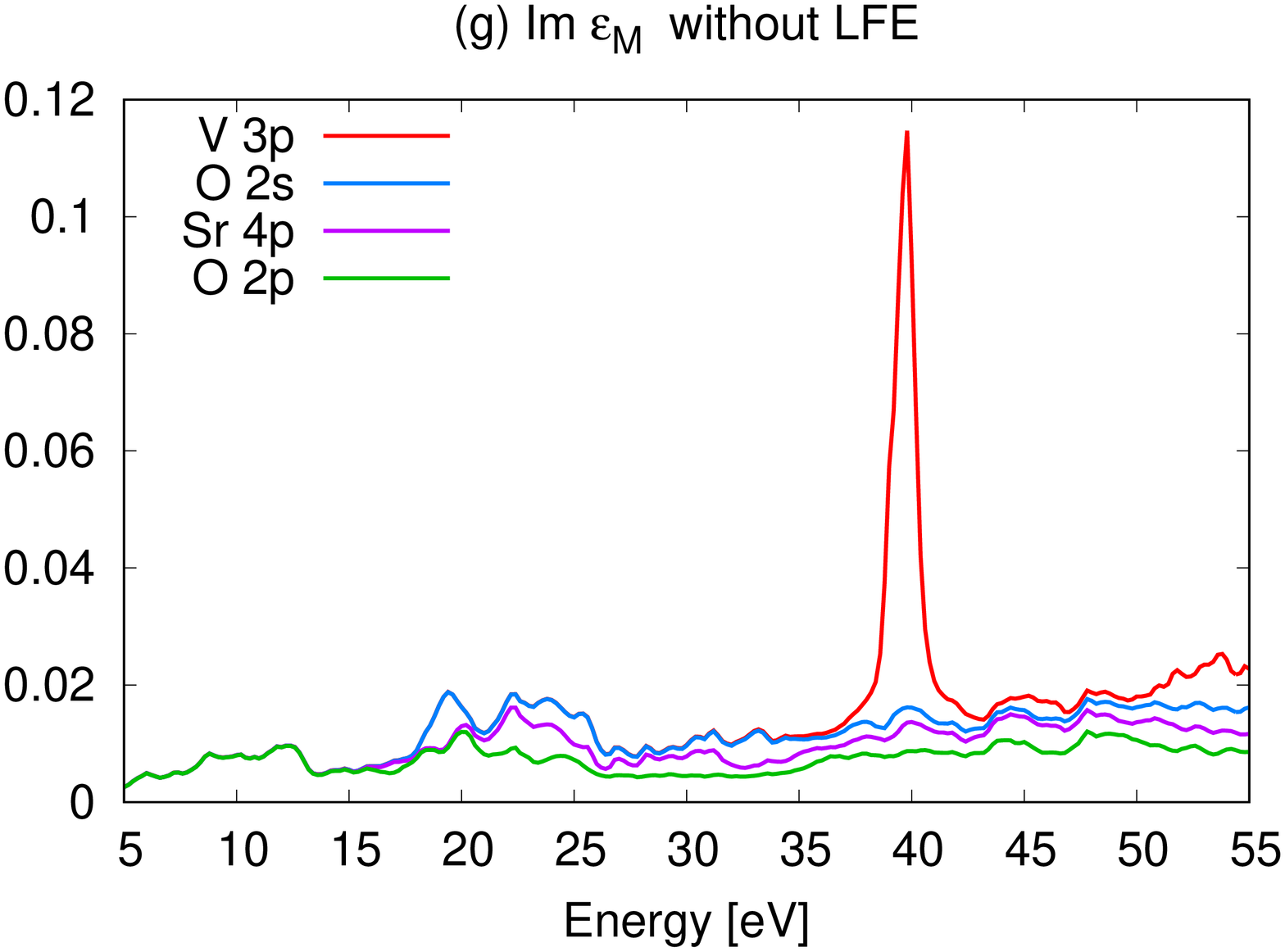}
\includegraphics[angle=270,width=0.32\textwidth,clip]{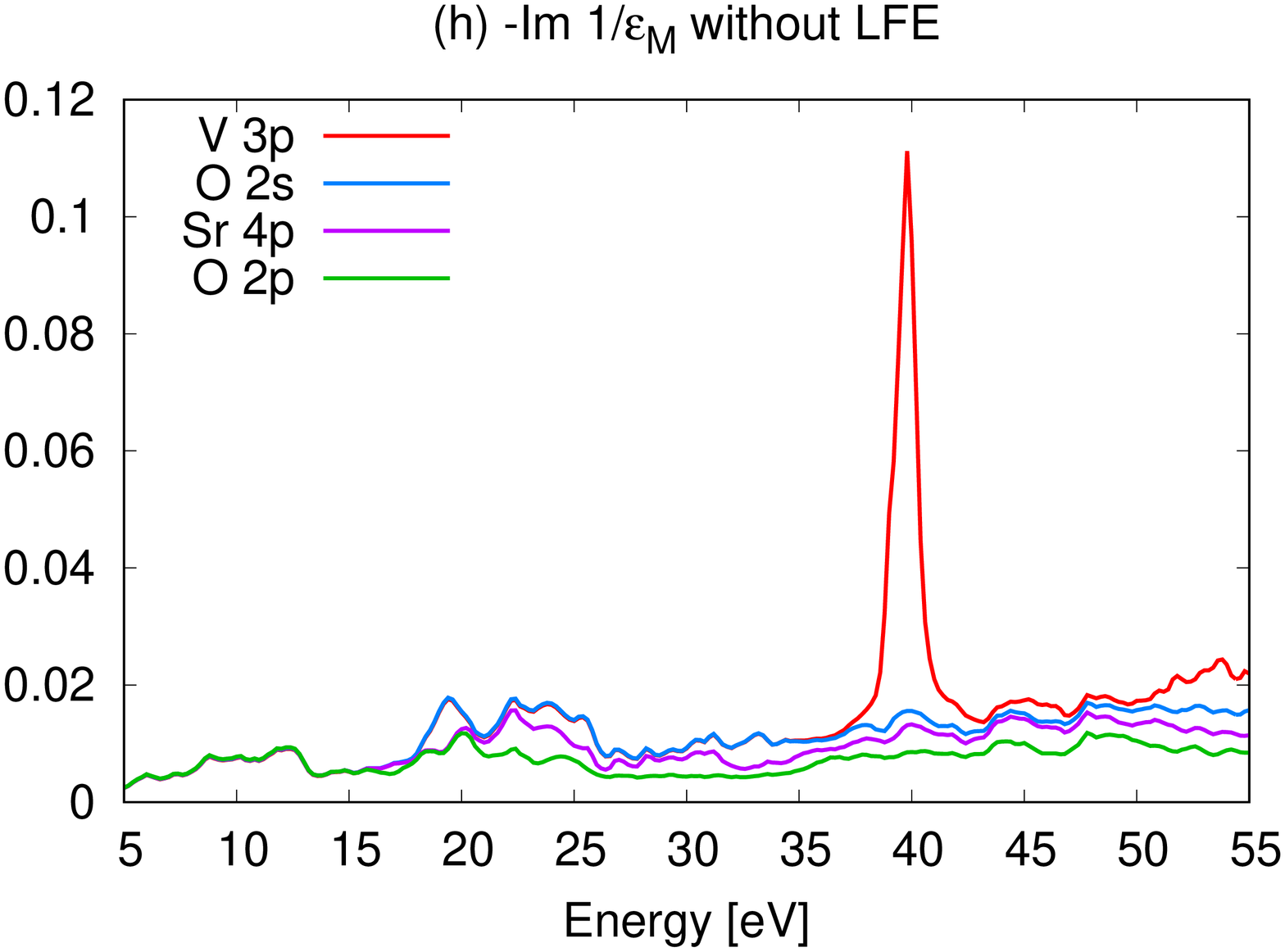}
\includegraphics[angle=270,width=0.32\textwidth,clip]{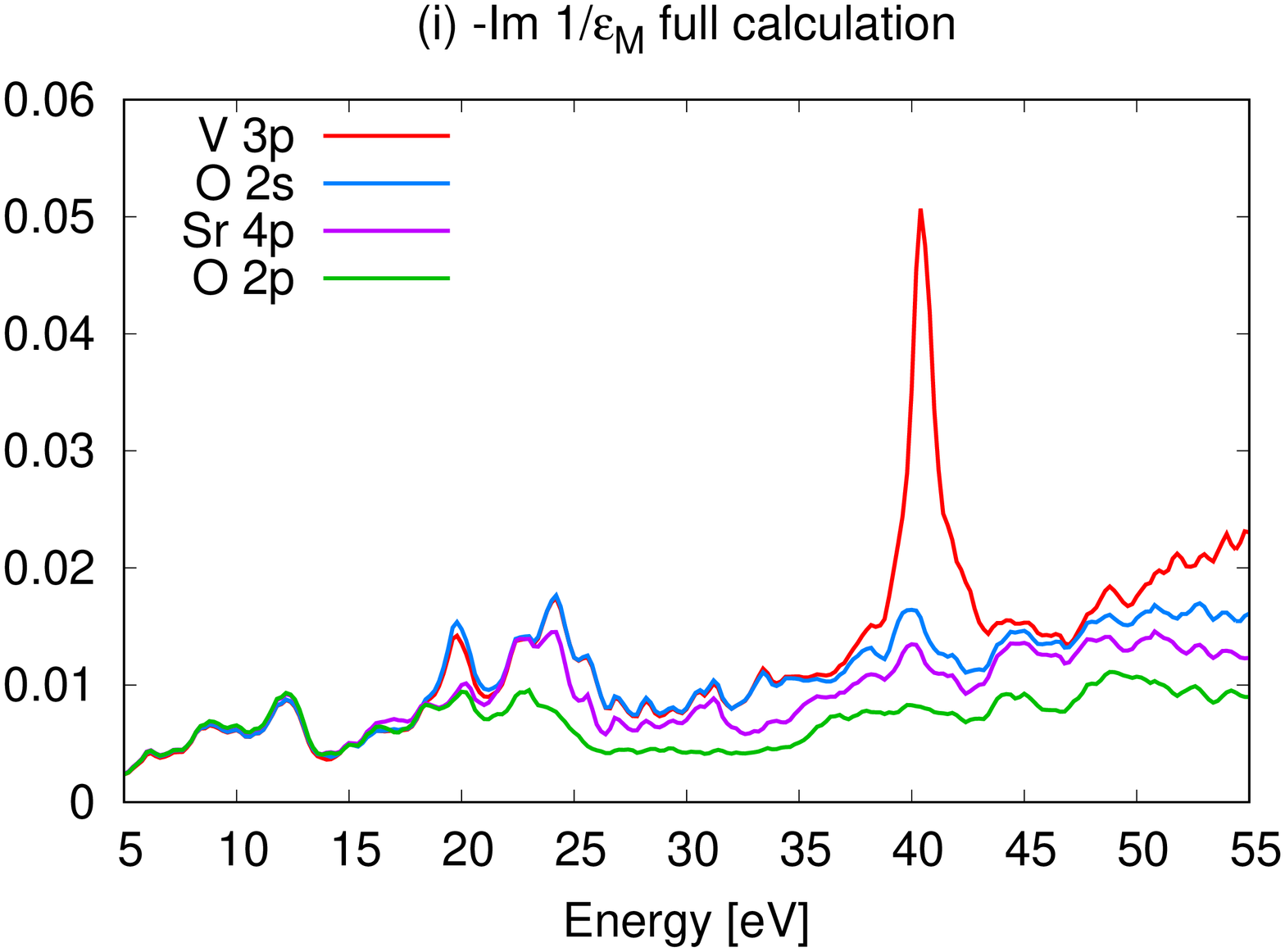}
\caption{Analysis of the contribution from O $2p$ valence electrons and different semicore states to the spectra: (Left) $\text{Im}\epsilon_M(\bfq,\w)$ in the independent-particle approximation (i.e., without LFE), see Eq. \eqref{epsnlf}, (center) the corresponding loss function $-\text{Im}\epsilon_M^{-1}(\bfq,\w)$, and (right) the full ALDA calculation for (upper row)  a small $q$=0.083, (middle row) an intermediate $q$=1.916, and (bottom row) a high $q$=3.916 for momentum transfers $\bfq$ along the (100) direction.} \label{fig3}
\end{figure*}

In the following we will analyse more in detail the physical origin of the most important peaks in the spectra and we will elucidate further the role of LFE.

\subsection{Origin of the spectral features}

In the calculations it is possible to examine the contribution to the spectra from the various one-particle  states by selectively removing them from (or adding them to) the sum in Eq. \eqref{eq:chi0}. To this end, we choose 3 representative spectra from the series displayed in Fig. \ref{fig2} for momentum transfers $\bfq$ along the (100) direction.
The outcome of this analysis is shown in Fig. \ref{fig3}(a)-(c) for the small $q =(1/12)G$, in Fig. \ref{fig3}(d)-(f) for the intermediate $q = (23/12) G$, and in Fig. \ref{fig3}(g)-(i) for the large $q = (47/12) G$.
In these figures we progressively add to the contribution from V $t_{2g}$ and O $2p$ valence bands (green lines) the shallow core (i.e. semicore) states\footnote{We note that O $2s$ and Sr $4p$ are also partially hybridized.}: Sr $4p$ (purple lines), O $2s$ (blue lines) and V $3p$ (red lines). The latter  corresponds to the full calculation, as V $3s$ do not contribute in the considered energy range and  Sr $4s$ (not shown) are weakly visible only at large $q$ (Sr $4s \rightarrow 4d$ is dipole forbidden).

Fig. \ref{fig3}(a) displays $\text{Im}\epsilon_M$ calculated in an independent-particle approximation (i.e., within the RPA without LFE), see Eq. \eqref{epsnlf}.
At energies below 15 eV the excitations from the O $2p$ dominate, while the peak at 20-25 eV is mainly associated to Sr $4p \rightarrow 4d$ transitions, with a very small contribution from the O $2s$. Finally the peak at 40 eV has a clear V $3p$ origin.

In the corresponding loss function  $-\text{Im}\epsilon_M^{-1}$, see Fig. \ref{fig3}(b), the peaks are shifted to higher energies. The difference between the absorption spectrum and the loss function can be understood in terms of 
 the long-range Coulomb interaction \cite{Onida2002,Sottile2005}.  The peaks at 13  eV and 29 eV in the calculation with all the states [red line in Fig. \ref{fig3}(b)] are plasmons: they are associated to zeros of $\text{Re}\epsilon_M$ (not shown). The structure at 40-44 eV is easily identified as  V $3p$ interband transition matching well the corresponding peak of $\text{Im}\epsilon_M$. The semicore V $3p$ peak also partially screens the main plasmon at smaller energy. This can be understood by comparing the full calculation [red line in Fig. \ref{fig3}(b)] with the one where V $3p$ are excluded (blue line). 
 
 It is noteworthy to remark that also the Sr $4p$ (purple line) give an important contribution to the main valence plasmon at 29 eV, whereas the collective charge excitation resulting from O $2p$ valence electrons only (green line) would have a much smaller energy. 
This result, as found also in other complex oxides\cite{Huotari2010}, clearly indicates that the valence charge dynamics of SrVO$_3$ cannot be described accurately by considering  only  the  V $3d$ and  O $2p$ electrons, but  also the  Sr $4p$ and  O $2s$ should  be  taken  into  account.  Indeed, with only  V $3d$ and  O $2p$ electrons (amounting to 19 electrons per unit cell), one would obtain 
a classical plasma frequency of only $\sim$21 eV for $\bfq\rightarrow0$ instead of 29 eV. 
 We can also compare SrVO$_3$ with metallic VO$_2$, which shares the same $d^1$ electronic configuration.
 While they have similar low-energy excitations\cite{Gatti2013,Gatti2015b} (and also analogous V M$_{2,3}$ edges \cite{Abe1997,gatti_phd}), the main plasmon energies are different: in VO$_2$ it is only 25  eV\cite{Abe1997,gatti_phd}. 
This difference points again towards the active role of shallow Sr core states that are obviously present only in SrVO$_3$. Indeed, the main plasmon energy of SrVO$_3$  bears a closer resemblance with SrTiO$_3$, another cubic perovskite that is a $d^0$ insulator, where the  plasmon is at 30 eV \cite{Ruotsalainen2015}.

Besides blueshifting by less than
1 eV the two plasmon resonances, the main effect of crystal local fields on the loss function, calculated within the ALDA in Fig. \ref{fig3}(c), is to suppress the semicore V $3p$ peak, which becomes a weak structure centred at 50 eV. Consequently the impact of V $3p$ on the plasmon is also strongly reduced\cite{Cudazzo2014}.  
We find here the typical depolarization effect due to the crystal local fields: the external perturbation creates additional microscopic Hartree potentials that act as counteracting local fields and lead to a weakening of the effective perturbing potential. The interband transitions involving semicore states correspond to charge excitations that are at the same time highly polarizable and localized. Both aspects concur to strong LFE\cite{Aryasetiawan1994,Vast2002,Cudazzo2014}.

We now examine what is happening at larger $\bfq$, where the role of the long-range Coulomb interaction is not crucial anymore \cite{Onida2002,Sottile2005}. As a consequence, also the loss function resembles $\text{Im}\epsilon_M$, compare Fig. \ref{fig3}(d) and Fig. \ref{fig3}(e), or Fig. \ref{fig3}(g) and Fig. \ref{fig3}(h). At this point the plasmons are completely damped by the coupling with electron-hole excitations (Landau damping). In the loss spectra they are replaced as most prominent features by the interband transitions that are the main peaks of $\text{Im}\epsilon_M$.
The largest variations for the $\text{Im}\epsilon_M$ spectrum with respect to the small $\bfq$ case in  Fig. \ref{fig3}(a) are the progressive decrease of the oscillator strengths of the O $2p$ contribution and the corresponding relative increase of the V $3p$ peak with respect to the Sr $4p$. At the largest $q$,  Fig. \ref{fig3}(g)-(h), the spectra are dominated by the V $3p$ peak rising over a flat background given by all the other contributions.

Investigating the loss function at larger wavevectors $\bfq$ corresponds to probing electronic excitations that are localised in real space. The influence of microscopic local fields therefore increases. In the intermediate momentum range, compare Figs.  \ref{fig3}(e) and \ref{fig3}(f), LFE change completely the spectral shape, as the result of the mixing of many transitions. Even at the largest $\bfq$, where only one $3p$ prominent peak remains, LFE quantitatively affect the spectra, strongly reducing the intensity of the V $3p$ peak with respect to the background [see Figs.  \ref{fig3}(h) and \ref{fig3}(i)].

The measurement by IXS of these semicore excitations, which is generally called x-ray  Raman  scattering\cite{Suzuki1970,Schulke2007,Huotari2017} (XRS), provides the same information as extreme ultraviolet absorption spectroscopy on the same shallow core edges 
but with the advantage of the possibility of tuning the momentum transfer, possibly disclosing nondipolar transitions. Here we conclude that for both XRS and absorption spectroscopies the strong LFE invalidate the simple picture of independent electron-hole transitions [see Eq. \eqref{epsnlf}]. In order to obtain a quantitative agreement with experiment (see Fig. \ref{fig1}), going beyond the independent-particle picture and considering the mixing of electron-hole transitions through LFE turn out to be crucial.

\subsection{Crystal local fields beyond depolarization effects}

Considering Eq. \eqref{lfe}, LFE can be understood as the result of the coupling of excitations with wavevectors $\bfqr$ and $\bfqr+\bfG$ belonging to different Brillouin zones identified by the reciprocal lattice vectors $\bfG$. As discussed in the previous sections, commonly LFE induce a damping and a blueshift of the peaks in the spectra, a mechanism that is physically associated to depolarization effects. It is hence striking when LFE instead make new features appear in the spectra. Here we will examine two of such cases in SrVO$_3$.

\begin{figure}[t]
\includegraphics[angle=270,width=\columnwidth,clip]{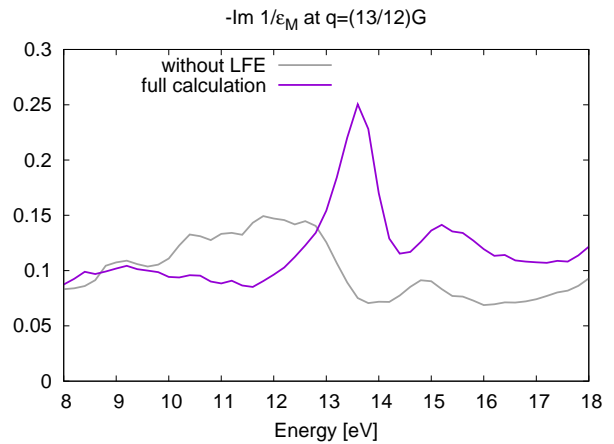}
\caption{Loss function at q=(13/12)$|\bfG|$ along the (100) direction calculated with or without LFE.} 
\label{fig4}
\end{figure}

The first feature is the peak at $\sim$ 13 eV  -- see Fig. \ref{fig2}(a) --
that displays an intensity dependence on $q$ that is remarkably nonmonotonic  and showing a maximum in the intermediate $q$ range, i.e., qualitatively different from the standard behavior  in  simple  materials. 
A similar finding was similarly reported for TiO$_2$ in Ref. \onlinecite{Gurtubay2004}. 
In both cases, LFE are responsible for the presence of this peak, which is here confirmed also by our experiment (see Fig. \ref{fig1}).
In Fig. \ref{fig4} we examine the loss function for $q=13/12$ calculated with and without LFE. We see that when LFE are neglected the loss function is featureless, whereas the inclusion of LFE induces the appearance of a sharp peak, which is located at the same energy of the plasmon at $q=1/12$ [see red lines in Fig. \ref{fig3}(b)-(c)]. Therefore we can conclude that LFE are the key for the coupling between independent electron-hole excitations at large $q$ and the first-Brillouin-zone plasmon. This effect is well known in layered materials \cite{Cai2006,Balassis2008,Silkin2009,Hambach2008,ralf_phd,Cudazzo2017,Echeverry2012,Cudazzo2012,Cudazzo2014b,Faraggi2012},
which are intrinsically inhomogeneous in the direction perpendicular to the layers, and where LFE give rise to 
a long-lived plasmon that continues to exist well beyond the first Brillouin zone. In  these situations, the mechanism behind the reappearance of the first-Brillouin-zone $\bfqr$ spectra at larger momentum transfers  $\bfqr+\bfG$ can be understood within a two-plasmon-band model\cite{Oliveira1980,ralf_phd}.

\begin{figure}[t]
\includegraphics[angle=270,width=\columnwidth,clip]{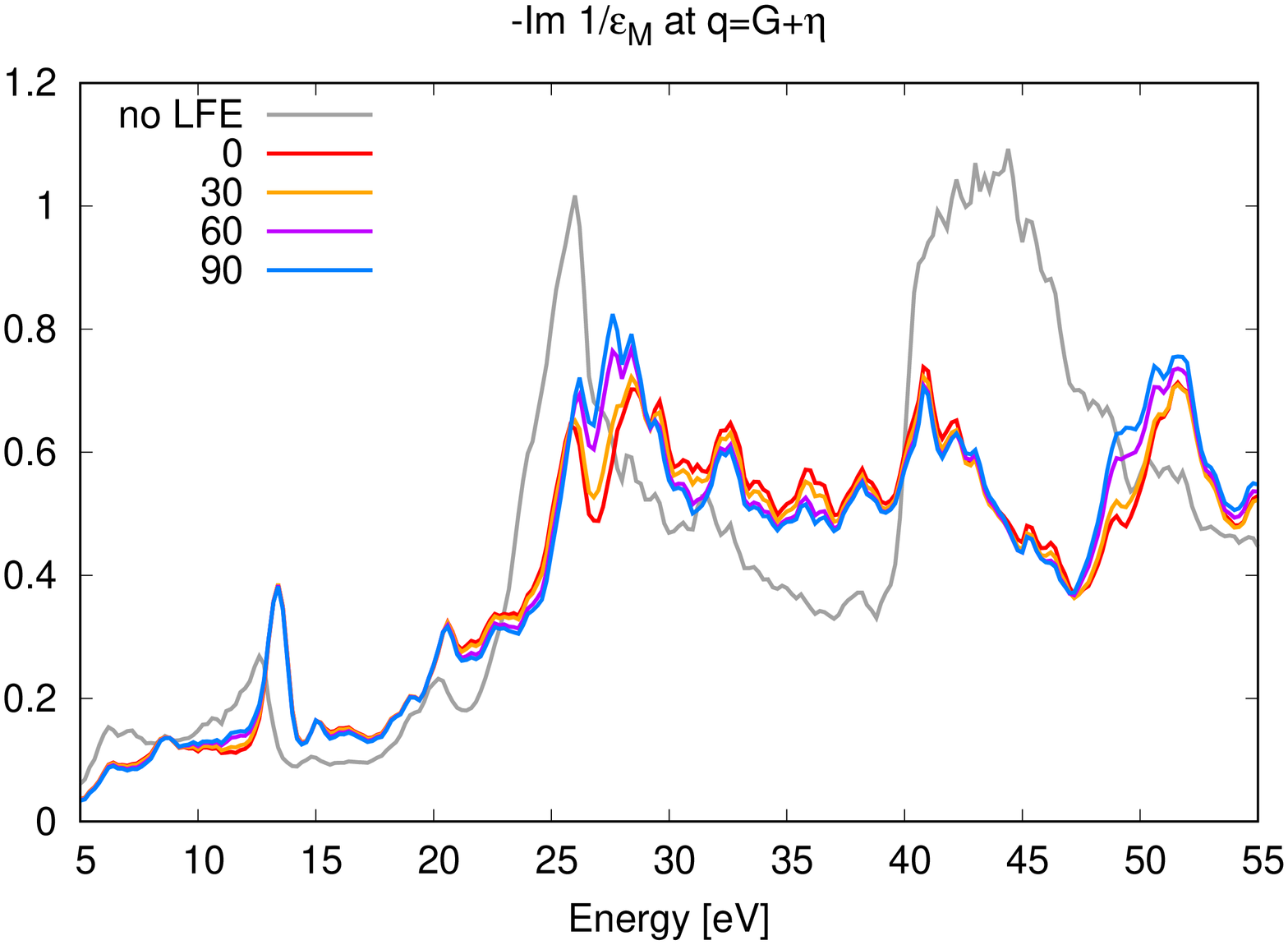}
\caption{Loss functions calculated at $\bfq=\bfG+\bm{\eta}$ where $\bfG=(100)$ and $\bm{\eta}$ has a very small size $|\eta|=10^{-3}$ and forms an angle $\alpha$ with  $\bfG$ that varies  from $0{\degree}$  when $\bm{\eta}$ is along $(100)$ to  $90{\degree}$ when $\bm{\eta}$ is along $(010)$. The spectra calculated without LFE (grey curve) are independent of $\alpha$, while spectra that include LFE are sensitive to the angle $\alpha$ with the small  $\eta$ component.} 
\label{fig5}
\end{figure}

The same mechanism also explains the angular anomaly (firstly discovered by Hambach {\it et al.} in graphite\cite{Hambach2008,ralf_phd}) that is observed here for the loss function calculated at  $\bfq = \bfG + \bm{\eta}$ where $\bfG=(100)$ and $\bm{\eta}$ has very small size $\eta=10^{-3}$ and direction identified by the angle   $\alpha$ between $\bfG$ and $\bm{\eta}$ that varies  from $0{\degree}$  when $\bm{\eta}$ is along $(100)$ to  $90{\degree}$ when $\bm{\eta}$ is along $(010)$: the loss function is found to be strongly dependent on the small $\bm{\eta}$, see Fig. \ref{fig5}. Neglecting LFE, instead, all the calculated spectra remain the same (see grey line in Fig. \ref{fig5}).
At first sight, this result seems even more surprising in the present case than in graphite, since SrVO$_3$ is a cubic material. However, a similar behavior has been also detected in silicon\cite{ralf_phd}, where it is also responsible for the presence of a plasmon-Fano resonance in the spectra \cite{Sturm1992}. 

By supposing that the main coupling occurs between $\bfqr$ inside the first Brillouin zone (i.e., for $\bfG=0$) and $\bfqr+\bfG$,   
Eq. \eqref{lfe} can be simplified to: 
\begin{multline}\label{eq2}
\epsilon^{-1}_{\mathbf{G},\mathbf{G}}(\mathbf{q}_r,\omega) = \frac{1}{\epsilon_{\mathbf{G},\mathbf{G}}(\mathbf{q}_r,\omega)}+  \\ \frac{\epsilon_{\mathbf{G},\mathbf{0}}(\mathbf{q}_r,\omega)\epsilon_{\mathbf{0},\mathbf{G}}(\mathbf{q}_r,\omega)}{[\epsilon_{\mathbf{G},\mathbf{G}}(\mathbf{q}_r,\omega)]^2}\epsilon^{-1}_{\mathbf{0},\mathbf{0}}(\mathbf{q}_r,\omega).
\end{multline}
The first term in Eq. \eqref{eq2} gives rise to the loss function at $\bfqr+\bfG$ obtained neglecting LFE (grey lines in Figs. \ref{fig4} and \ref{fig5}). 
The second term accounts for the LFE, showing explicitly the coupling with the first-Brillouin-zone dielectric function  $\epsilon^{-1}_{\mathbf{0},\mathbf{0}}$, weighted by its off-diagonal elements. When LFE are not negligible -- as in the case of SrVO$_3$  -- this term explains both the reappearance of the small $\bfqr$ plasmon and the angular anomaly.
When the weighting factor is not much frequency dependent (and the other off-diagonal terms $\mathbf{G}' \neq \mathbf{G}$, here neglected, are not important), from Eq. \eqref{eq2} we see that the spectrum at $\bfqr+\bfG$ indeed results being proportional to the spectrum at $\bfqr$. 
Moreover, the angular    anisotropy in the spectra at $\bfG + \bm{\eta}$ analogously stems from the off-diagonal elements $\epsilon_{\mathbf{G},\mathbf{0}}(\bm{\eta})$ and $\epsilon_{\mathbf{0},\mathbf{G}}(\bm{\eta})$. Indeed both depend on $\bm{\eta}$, while the first term in Eq. \eqref{eq2} does not (see grey line in Fig. \ref{fig5}; since SrVO$_3$ is cubic, also $\epsilon^{-1}_{\mathbf{0},\mathbf{0}}(\bm{\eta})$ is independent of $\bm{\eta}$). 

In summary, the coupling between spectra from different Brillouin zones associated to the LFE explains the strong dependence of the spectra on the small $\bm{\eta}$ component, which  is in contrast to what one would naively expect \footnote{It is also interesting to note that, contrary to graphite \cite{Hambach2008,ralf_phd}, in this case LFE apparently result also from the coupling with other $\mathbf{G}'\neq\mathbf{G}$ elements. Following Hambach \cite{ralf_phd}, for a more quantitative analysis one should hence use a two-plasmon-band model.}.

\section{Conclusions}
\label{conclusions}

We have characterised the dynamical screening properties of SrVO$_3$ by comparing experimental IXS spectra and first-principles TDDFT calculations. Our results provide a useful reference for both DMFT and GW calculations.
The very good agreement between experiment and theory has allowed us to analyse in detail the physical origin of the main electronic excitations in the 5-50 eV energy range,
together with their dispersion as a function of the wavevector $\bfq$.
Our investigation leads us to conclude that crystal local field effects are an essential feature of the dielectric response of correlated materials that should not be overlooked in future studies. 

\acknowledgements

This work has been supported by the Labex PALM (Grant No.  ANR-10-LABX-0039-PALM).
Computational time was granted by GENCI (Project  No.  544). We acknowledge the European Synchrotron Radiation Facility (ESRF) for providing synchrotron radiation and Christian Henriquet for technical support during the beamtime. DP would like to acknowledge support from the Engineering and Physical Sciences Research Council (EPSRC-EP/T028637/1) and support from the Oxford-ShanghaiTech collaboration project

\bibliographystyle{apsrev4-1}
%

\end{document}